\documentclass{article}
\usepackage{spconf,amsmath,graphicx, booktabs}
\usepackage{pgfplots}
\usepackage{enumitem}
\pgfplotsset{grid style={dashed,gray}}
\pgfplotsset{compat=1.12}
\usepackage{tikz}
\usetikzlibrary{shapes.geometric}
\usetikzlibrary{shapes.misc,shadows}
\usetikzlibrary{positioning} 
\usetikzlibrary{arrows} 
\usetikzlibrary{arrows.meta}
\tikzset{%
    >={Latex[width=1mm,length=1mm]},
    base/.style = {
        rectangle, rounded corners, draw=black,
        minimum width=2cm, minimum height=.4cm,
        text centered, font=\tiny},
    acoustic_model/.style = {base, fill=red!15},
    language_model/.style = {base, fill=cyan!20},
    joint/.style = {base, fill=yellow!15},
    io/.style = {base, fill=none, draw=none, minimum width=0cm},
    data/.style = {
        rectangle, draw, align=center, left color=blue!20, right color=white, 
        minimum width=0.5cm, minimum height=0.5cm},
    block/.style ={
        rectangle, thick, draw=black, align=center, fill=orange!15,
        minimum height=.3cm, minimum width=2cm, text width=2cm},
    connector/.style={-latex, font=\tiny},
    rectangle connector/.style={
        connector,
        to path={(\tikztostart) -- ++(#1,0pt) \tikztonodes |- (\tikztotarget) },
        pos=0.5
    },
}
\usepackage{color}

\newif\ifblind
\blindfalse

\title{Improved long-form speech recognition by jointly modeling the primary and non-primary speakers}

\ifblind
    \name{BLIND}
    \address{BLIND}
\else
    \name{\begin{tabular}{cc}
Guru Prakash Arumugam, Shuo-yiin Chang, Tara N. Sainath\\Rohit Prabhavalkar, Quan Wang, Shaan Bijwadia
\end{tabular}}
    \address{Google LLC, Mountain View, CA, U.S.A. \\
    \{guruprakash, shuoyiin, tsainath\}@google.com}
\fi

\copyrightnotice{979-8-3503-0689-7/23/\$31.00~\copyright2023 IEEE}
\begin{document}
%
\maketitle

\begin{abstract}
ASR models often suffer from a long-form deletion problem where the model predicts sequential blanks instead of words when transcribing a lengthy audio (in the order of minutes or hours). From the perspective of a user or downstream system consuming the ASR results, this behavior can be perceived as the model ``being stuck", and potentially make the product hard to use. One of the culprits for long-form deletion is training-test data mismatch, which can happen even when the model is trained on diverse and large-scale data collected from multiple application domains. In this work, we introduce a novel technique to simultaneously model different groups of speakers in the audio along with the standard transcript tokens. Speakers are grouped as primary and non-primary, which connects the application domains and significantly alleviates the long-form deletion problem. This improved model neither needs any additional training data nor incurs additional training or inference cost.   
\end{abstract}
\section{Introduction}
\label{section:intro}

End-to-end automatic speech recognition (ASR) systems ~\cite{pmlr-v32-graves14, chan2021speechstew, rao2017exploring} are capable of handling different scenarios (e.g. a few seconds to hours in length, clean audio vs noisy audio with background speech or music) and different downstream applications such as voice assistants (Google, Siri, Alexa, etc.) and captioning YouTube videos. These systems are trained with data from different sources and noise conditions for robust performance ~\cite{narayanan2018toward, sainath2020streaming}. One problem for such models is long-form deletion --- sometimes the model introduces high deletion errors when the input audio is long. In such cases, the user may perceive the system as being stuck (no words are emitted intermittently), which can significantly hurt the user retention rate of the end product. Alternatively, if the transcript from ASR is consumed by downstream systems for semantic understanding, the missing words can induce cascading errors.

There have been many efforts in the literature to understand the cause of long-form deletions and attempt to improve general long-form quality. Some studies indicate that training-test data mismatch (due to acoustics, noise, audio length differences) is one cause ~\cite{narayanan2019recognizing, chiu2019comparison, lu2021input}. The long-form deletion problems could be introduced in both models with recurrent structure (e.g. LSTM-based ASR ~\cite{narayanan2019recognizing, chiu2019comparison}) and non-recurrent structure (e.g. conformer based ASR ~\cite{lu2021input}).
~\cite{kim2018dialog, kim2019gated, hori2021advanced} attempts to use previous context as additional input to the model to improve long-form performance. ~\cite{zeyer2023monotonic} indicates that model architectures with global attention do not scale well to long-form audio and propose smaller attention windows for better scalability. But they do not explicitly target the long-form deletion problem directly. Although improved model architectures can potentially alleviate the problem, we still sometimes see that the problem exists. This paper looks at an alternative approach to fix it.

In this work, we demonstrate that even a model without global attention and trained with vast amount of diverse data can have significant long-form deletion problem. We create a targeted testset and metric that shows the long-form deletion clearly and analyze when the model exhibits the problem. We identify that a model trained on data from different applications can exhibit the long-form deletion problem for some domains more prominently than others. Different application domains expect different types of speakers to be transcribed. For example, for a voice assistant (e.g. Google, Siri), ASR is expected to only transcribe the target user talking to the device (i.e. the primary speaker) and ignore other speakers. However, when captioning a YouTube video, ASR is expected to transcribe all speakers. These expectations are reflected in the ground truth annotations of both training and test data.
If data from these domains are directly mixed together without special handling, the conflicting expectations from different domains will cause confusions during the model training when non-primary speakers are present --- to transcribe or not to transcribe, that is the question.
These conflicting training goals ultimately limit the model from learning cross-domain capabilities, thus leading to long-form deletion errors.

To address the above mentioned problem, we simultaneously model the speakers along with the ASR transcript tokens to create better synergy across domains. We propose to expand the ASR output vocabulary with two novel tokens: \texttt{<end-primary>} and \texttt{<end-others>}, and show how to train and run inference with ASR.
 
There are many approaches to model the speakers in an audio segment, such as speaker recognition, speaker diarization, speaker turn detection, speaker-attributed ASR, and speech separation. We will briefly review these work in Section~\ref{sec:related_work}. But in summary, our work differs from those approaches by explicitly identifying the primary speaker, transcribes audio in a single inference pass without additional steps. Our work also differs with the end goal being long-form deletion improvement as opposed to improving speaker detection or speaker turn accuracy.

We investigate the Word Error Rate (WER), endpointing latency (EP) on multiple testsets. Additionally, we measure the number of times 25 words are deleted in sequence on the targeted testset for long-form deletion. With the proposed technique, we show that the WER and EP are neutral compared to baseline and the long-form deletion metric improves by 55\%.

\section{Method}
\label{section:method}

\subsection{Modeling primariness of speakers}

A single end-to-end (E2E) ASR model serving different use cases is generally preferable to separate models, which could be more expensive to train, maintain and serve. Modern ASR systems are capable of training on large amounts of diverse data and serve multiple downstream applications e.g. ~\cite{sainath2020streaming}. One way to group these applications would be:

\begin{enumerate}[leftmargin=*]
\item \textbf{Short domain}: User making a short request (few seconds) with voice assistant (e.g. Alexa or Siri) or a short query in search bar (like Google voice search).
\item \textbf{Dictation domain}: User dictating a (potentially) long query (like composing an email or message by speaking instead of typing).
\item \textbf{Caption domain}: Generating captions (like for YouTube videos, live-streams).
\end{enumerate}

For Short and Dictation, there is one unique speaker who should be considered as the user and the ASR model is expected to transcribe only their speech. If there is any background noise (e.g. TV noise, music) or other speakers, the model should ignore them. For Caption, the goal is to transcribe all the speech (e.g. a podcast with multiple participants). We would like to serve all these domains (Short, Dictation and Caption) with a single E2E ASR model. However, the different expected output for different domains in terms of whose speech should be transcribed is a potential challenge for the model. We aim to simplify this by explicitly modeling the different speakers in a given audio segment. We group the speakers into \texttt{Primary} and \texttt{Non-Primary}. For Short and Dictation, \texttt{Primary} is the target user and \texttt{Non-Primary} is all other speakers / background noise. This distinction between primary and non-primary is necessary, since for Short and Dictation the ASR model is expected to not transcribe the non-primary speakers. The two groups can be modeled by the addition of two new tokens to the ASR output vocabulary: 

\begin{enumerate}[leftmargin=*]
    \item \texttt{<end-primary>} - Used to indicate when the primary speaker is done speaking. Either there is a speaker turn after this event and other speakers begin speaking; or this is the end of the audio.
    \item \texttt{<end-others>} - Used to indicate when all other speakers are done speaking. Either there is a speaker turn after this event and primary speaker speaks again; or this is the end of the audio.
\end{enumerate}

\begin{figure}[t]
  \centering
  \includegraphics[scale=0.2]{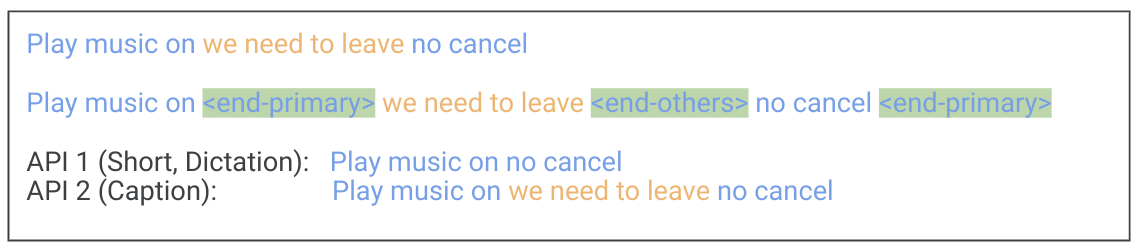}
  \caption{Primary speaker text in blue, others text in yellow}
  \label{fig:sptags-api}
\end{figure}

For instance, if the primary speaker says ``Play music on ... no cancel" and another background speaker says ``but we need to leave" during the utterance, the expected ASR output would be ``Play music on \texttt{<end-primary>} but we need to leave \texttt{<end-others>} no cancel \texttt{<end-primary>}". With this protocol, two APIs can be exposed to serve all application domains. For Short and Dictation domains, the ASR output (containing all speakers transcript and speaker-tag tokens) can be processed to remove non-primary speakers transcript. And for the Caption domain, the ASR output can be processed to contain all speakers' transcript. This is shown in Figure \ref{fig:sptags-api}. Now, the ASR output is uniform for all domains (Short, Dictation and Caption) i.e. transcribe all speakers and output the speaker-tag tokens appropriately.

We call these new tokens ``speaker-tags". Note that this does not explicitly handle overlapping speech by both primary and a non-primary speaker. This is a potential limitation that can be an area of research for future work. It's also worth noting that we can alternatively use \texttt{<begin-primary>} and \texttt{<begin-others>} to indicate when a speaker starts speaking (instead of finishes speaking). By definition, these would be equivalent to speaker-tags but we do not pursue this as emitting tokens earlier rather than later could be more difficult for the model.

\subsection{Relabeling training data with speaker-tags}
\label{sec:relabel}

To create training data in the above format, we introduce a technique that can relabel existing data without manual examination. The technique utilizes two teacher models:

\begin{enumerate}[leftmargin=*]
    \item 600M parameter bidirectional teacher model trained with Short and Dictation domain data only with the architecture and procedure described in ~\cite{hwang2022pseudo}. In particular, this model is trained only on supervised data from domains which only preserve the primary speaker transcript. We call this ``Teacher-Primary".
    \item 183.5M parameter bidirectional model trained on short-segments of YouTube data as described in ~\cite{chiu2021rnn} and further finetuned with 1k hours of supervised YouTube data. In particular, this model is trained on data from domains where all (primary, secondary, background noise) speech has to be transcribed. We call this ``Teacher-All".
\end{enumerate}

\begin{figure}[t]
  \centering
  \includegraphics[scale=0.19]{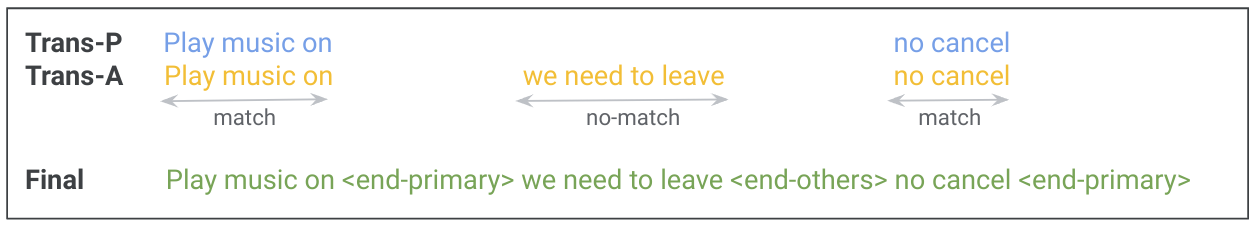}
  \caption{Technique based on sub-sequence match between transcripts. Trans-P is Trans-Primary and Trans-A is Trans-All}
  \label{fig:heuristic}
\end{figure}

For a given audio segment, suppose we know the transcript for just the primary speaker (say Trans-Primary) and the whole transcript (say Trans-All). We do a sub-sequence match between the two to add the two speaker tag tokens at appropriate boundaries as shown in Figure \ref{fig:heuristic}.

For Short and Dictation training data, the available ground truth transcript would be Trans-Primary. Trans-All can be obtained by running the Teacher-All. Similarly, for Caption data, the available ground truth transcript would be Trans-All. Trans-Primary can be obtained by running the Teacher-Primary. Although Caption audio data may not have a clear \texttt{primary} speaker, the primary speaker can be identified implicitly based on the matching. 

For this process to succeed, there must be exactly one sub-sequence match between Trans-Primary and Trans-All. This may not happen for several reasons:

\begin{enumerate}[leftmargin=*]
    \item Teacher models make mistakes and human-transcription could be incorrect or inconsistent resulting in no sub-sequence match. 
    \item Normalization mismatches due to capitalization, punctuation and related differences e.g. ``where`s the Eiffel-tower located" vs ``Where is the Eiffel tower located ?".
    \item There might be multiple valid sub-sequence matches. e.g. Trans-Primary is ``how tall is Barack Obama" and Trans-All is ``how tall is it is the end Barack Obama" - where the word ``is" can be matched in multiple ways leading to different output transcripts.
\end{enumerate}

For robust data relabeling, all punctuation and capitalization is denormalized before matching. Additionally, any one word differences are not considered as a mismatch (e.g. ``how tall is Barack Obama" and ``how tall is a Barack Obama"). For other cases with no or multiple matches after these adjustments, the original transcript truth is used directly without any speaker-tag tokens inserted. Additionally, some training data could be very long and is segmented into smaller chunks for convenience (like YT-T data in \cite{zhang2022bigssl}). This audio segmentation boundary is arbitrary and not a speaker-tag transition boundary. When relabeling such segmented data, the last speaker-tag token at the end will be removed (e.g. ``Welcome to the show \texttt{<end-primary>} thank you \texttt{<end-others>} how are \texttt{<end-primary>}" becomes ``Welcome to the show \texttt{<end-primary>} thank you \texttt{<end-others>} how are"). 

\subsection{Related work}
\label{sec:related_work}
There are several technologies in the literature to model multi-speaker / background noise scenarios, but they are not ideal here.
Speaker recognition~\cite{wan2018generalized,li2017deep,snyder2018x} requires target voice enrollment~\cite{wang2020version}, which needs a separate step before ASR and it is not feasible for all applications (e.g. not all users will enroll their voice prior to using the product). Speaker diarization~\cite{anguera2012speaker,park2022review,zhang2022odysseytutorial,wang2018speaker} and speaker-attributed ASR~\cite{kanda2021investigation} anonymously label speakers, but do not explicitly identify who is the primary speaker. Speaker turn detection ~\cite{xia2022turn,zhao2023augmenting} would mark the boundaries between two speaker's speech, however it is also insufficient to distinguish the primary speaker. Speech separation ~\cite{hershey2016deep} usually suffers from ``permutation-invariance" (correctly ordering speakers across segments) and high computational cost due to running ASR on multiple separated channels, in addition to not identifying the primary speaker. Our work will explicitly identify the primary speaker (needed for Short and Dictation domains) and does not need any additional step outside ASR.

The method proposed here with introducing new tokens for separating speakers has some parallels to other work e.g. ~\cite{masumura2021unified, kanda2020serialized}. However, our work differs on a few fronts: how speakers are grouped into primary vs non-primary and our particular focus on long-form deletion problem.

\subsection{RNN-T ASR and speaker-tags}

\begin{figure}[!ht]
\centering
\hspace{-0.05in}
    \includegraphics[scale=0.27]{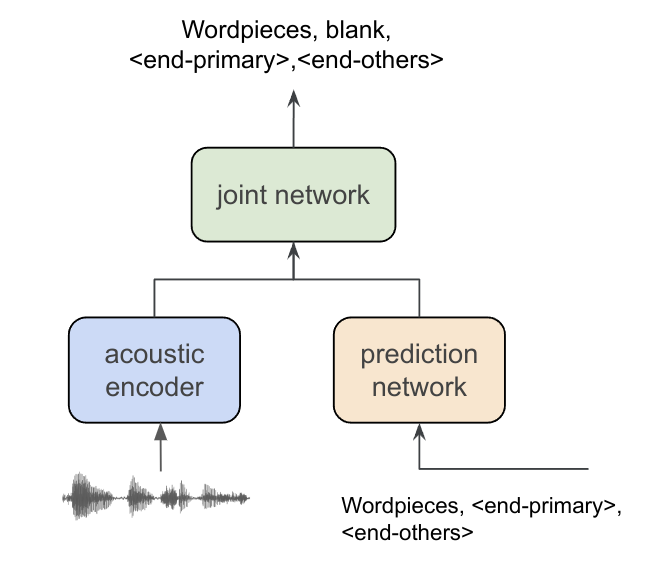}
    \vspace{-0.1in}
    \caption{E2E model architecture}
    \vspace{-0.15in}
    \label{fig:system_e2e}
\end{figure}

As shown in Fig. \ref{fig:system_e2e}, our E2E ASR models have the RNN-T architecture, where the label sequence can be augmented with the speaker-tag tokens.
At each time step $t$, the model would receive a new acoustic frame $x_t$ and output a probability distribution over \sloppy $y_t \in V \cup \{ \texttt{<end-primary>} , \texttt{<end-others>}\}$, where $V$ being the wordpiece vocabulary and a blank symbol.

The RNN-T network consists of three components: encoder, prediction network and joint network. The encoder is analogous to an acoustic model and the prediction network is analogous to a language model. At each time step, the encoder receives acoustic features and outputs acoustic encodings. The recurrent prediction network obtains the embedding vectors from $N$ previous (non-blank) tokens and produces the text encodings. The encodings from the encoder and prediction network are then fed into the joint network, which fuses the acoustic and text encodings together to compute the output logits. For the baseline model, the prediction network is conditioned only on the wordpiece tokens. But for the new proposed model, the prediction network is conditioned on the speaker-tag tokens as well.

During inference, we perform beam search decoding over the wordpiece tokens and the speaker-tag tokens:
\begin{equation}
y* = \operatorname*{arg\,max}_y \log P_{\text{asr}}(y|x_{t-k}, \cdots, x_{t}, y_{u-N}, \ldots, y_{u})
	\label{eqn:beam-search}
\end{equation}
where $\mathbf{x}_{t-k},... \mathbf{x}_t$ represents acoustic observations received by the encoder; $\mathbf{y}_{u-N},... \mathbf{y}_u$ stands for the sequence of wordpieces and speaker-tags; $k$ and $N$ are the context window sizes for the encoder and prediction network, respectively. By the definition of speaker-tags, \sloppy the \texttt{<end-primary>} and \texttt{<end-others>} should come in alternate order. But as per Eq. \ref{eqn:beam-search}, this is not explicitly enforced. It's possible that the model emits the same speaker-tag token repeatedly e.g. ``why is the \texttt{<end-primary>} sky blue \texttt{<end-primary>} welcome home \texttt{<end-others>}". But during post processing, each segment preceding the \texttt{<end-primary>} token can be added as the primary speaker transcript i.e. the above example is equivalent to ``why is the sky blue \texttt{<end-primary>} welcome home \texttt{<end-others>}". It's possible to enforce alternating the two speaker-tag tokens during beam search. But we do not purse that option as the above post processing is sufficient. Note that the training data will not have such repeated tokens.

\section{Experimental Setup}

\subsection{Datasets and Metrics}

The training data consists a mix of three application domains: Short, Dictation and Caption. The Short training data has about 500M utterances, totalling 500k hours. The Dictation set has about 10M utterances, totalling 23k hours. The Caption training data has about 250M utterances, totalling 600k hours. Dictation training data is small in scale compared to Caption and Short. Additionally, Dictation data is generally clean but Caption data is noisy (music, background speech or noise, multiple-speakers). All the data is either human-transcribed or machine-transcribed for semi-supervised training ~\cite{hwang2022pseudo, chiu2021rnn}. Multi-style training (MTR) \cite{kim2017generation} is also used for increasing the data diversity. For experiments with speaker-tags, the transcripts are modified as described in Section ~\ref{sec:relabel}. 

For evaluation, one test set is used for each domain. The Short testset has 9k utterances, totalling 12 hours. The Dictation testset has 17k utterances, totalling 39 hours. The Caption testset has 20k utterances, totalling 380 hours. On these testsets, the Word Error Rate (WER) is measured. The endpointer median latency (EP50) and 90th percentile latency (EP90) are also measured for the Short testset. Endpointer latency is defined as the time it takes for the microphone to close after the user finishes speaking as determined by forced alignment ~\cite{maas2018combining, chang2019joint}. This measurement is only required for Short domain since Dictation and Caption domains can have arbitrarily long audio input (order of minutes / hours) and the end of the audio is typically decided by the user or the end of the video file or other systems. A special \texttt{<end-of-speech>} token is included in the wordpiece vocabulary for endpointer.

\subsection{Long-form deletion testset and metric}
\label{section:measure-stuckiness}
\begin{figure}[t]
  \centering
  \includegraphics[scale=0.45]{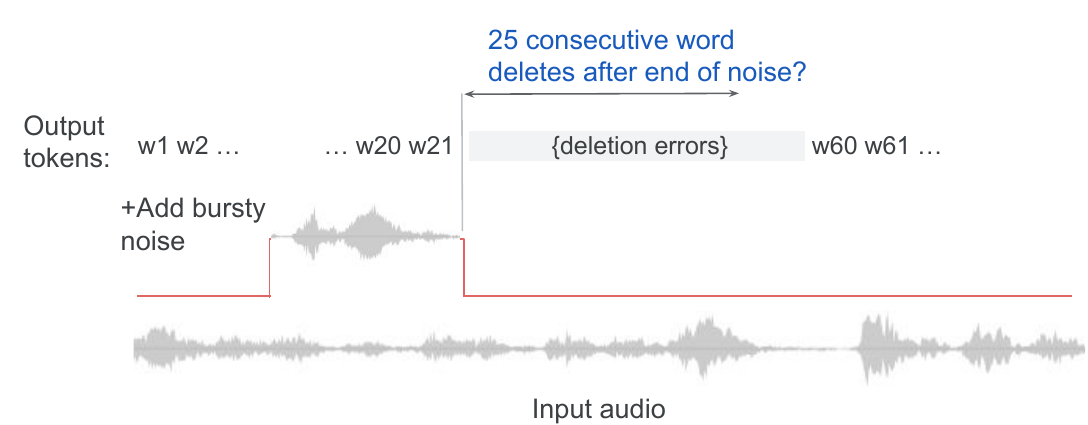}
  \caption{We measure long-form deletions by counting consecutive deleted words after a bursty noise.}
  \label{fig:stuckiness-diag}
\end{figure}

To the best of our knowledge, no existing literate proposes a dedicated testset or metrics for measuring the long-form deletion problem directly. Typically, Deletion WER is measured on a generic testset as a proxy, but this does not explicitly check for continuous deletions in sequence.  One contribution here is to build a novel testset and metrics to measure the long-form deletion problem reliably.
From user feedback and bug reports, the long-form deletion problem seem to occur prominently when there is static or background noise in the audio. Based on this observation, we create a testset by augmenting hand-transcribed long audio segments with bursty noise using approaches described  in~\cite{kim2017generation}. Bursty noise is a type of noise that only spans a finite time range in the entire audio, like a step function. The modified audio is transcribed by an ASR model and we check for long sequential deletion errors after the end of noise. The process is also shown in Figure \ref{fig:stuckiness-diag}.  

For this process, we choose to check for deletions after noise disappears since with noise it might be harder to separate the speech. This could be a different problem altogether e.g.~\cite{o2021conformer}. For this reason, we check for errors outside the noise range. Only variants with 25 sequential word deletions after the end of the noise are chosen for the testset.

After this process, the testset consists of 70 long audio queries totalling 20 hours. For this testset, the WER and the number of times 25 long sequential deletions (i.e. 25 words are sequentially deleted by the model) are measured. 

\subsection{Model description}

Experiments are run with a Conformer transducer model ~\cite{gulati2020conformer}. The encoder is based on the cascaded architecture ~\cite{narayanan2021cascaded, sainath2022improving} with 7 causal Conformer blocks without future context followed by 10 non-causal Conformer blocks with 900 ms of future acoustic context. The first two causal blocks do not have multi-head self-attention (MHSA) modules. The model dimensions are 512 and 640 for causal and non-causal encoders. All Conformer blocks have 8 attention heads in the MHSA modules. Convolution kernel size in Conv modules is 15; FFN module dimension is 4x the model dimension. In total, the causal encoder has 47M parameters and the non-causal encoder has 99M parameters. The output of both encoders are projected to 384 dimensions. The joint network combines features from the encoder and embedding from prediction network ~\cite{botros2021tied} with 640 units. Each decoder has 9.5M parameters. During training, the encoders are co-trained with equal weights. The model has FastEmit loss ~\cite{yu2021fastemit} to encourage lower user perceived latency and better endpointer performance. The output vocabulary has 4096 wordpieces and tokens, two of which are speaker-tag tokens. 

Models are trained with the Lingvo toolkit ~\cite{shen2019lingvo} for 450k steps on 8x8 tensor processing units (TPUs) with a batch size of 4096.

\section{Experiments and Results}

The E2E ASR model is trained at a large scale with diverse data from different application domains and is serving different use cases. A baseline model without speaker-tags can potentially make mistakes since the domains have different expectations (Short needs endpointer prediction; Short and Dictation expects only primary speaker transcription vs Caption expects all speech transcribed). To improve performance, a domain-id encoded by a 4-dimensional one-hot vector can be appended to the acoustic features ~\cite{sainath2020streaming}. The domain-id distinguishes Short vs Dictation vs Caption domains. As shown in Table \ref{tab:dom-id}, baseline model B1 (with domain-id input) outperforms the baseline model B0 (without domain-id input) on WER measurement significantly. Note that the results for B1 are with each domain correspondingly. The Short testset is evaluated with Short domain-id; the Dictation testset is evaluated with Dictation domain-id; and the Caption testset is evaluated with Caption domain-id. This provides the best performance and using any other combination (e.g. Caption domain-id for the Short testset) leads to much worse performance (over 20\% regression) and is not feasible.

\begin{table}
    \centering
    \small
    \caption{WER of different models on Short, Dictation (Dt.) and Caption (Cp.) testsets. Results are without endpointer.}
    \begin{tabular}{c c c c c}
        \toprule
        \textbf{Exp Id} & \textbf{Model} & \textbf{Short} & \textbf{Dt.} & \textbf{Cp.} \\  
        \midrule
            B0 & Base no domain-id input & 5.6 & 4.2 & 16.9 \\
            B1 & Base with domain-id input & 5.6 & 3.6 & 15.0 \\
            E1 & speaker-tags model & 5.6 & 3.6 & 15.3 \\
        \bottomrule
    \end{tabular}
    \label{tab:dom-id}
\end{table}

Table \ref{tab:del-res} however shows that with the addition of domain-id input, the model exhibits huge long-form deletions with the dictation domain-id. The same model evaluated with Caption domain-id has much better performance. This indicates that domain-id input can limit the cross-domain learning from training data. Caption training data is very rich (with different noise conditions, music, speakers etc) unlike Dictation training data. What the model learned from Caption training data does not translate well for Dictation domain-id causing worse long-form deletion. Note that the Short domain only has short audio (few seconds) and measuring long-form deletion would be meaningless. 

\begin{table}
    \centering
    \small
    \caption{Long-form deletion results. We report WER together with deletion / insertion / substitution (D/I/S). We also report the number of times 25 consecutive words are deleted (\# of 25 del.).}
    \begin{tabular}{c c c c}
        \toprule
        \textbf{Exp} & \textbf{Setup} & \textbf{WER (D/I/S)} & \textbf{\# of} \\ 
        \textbf{Id} & & & \textbf{25 del.} \\  
        \midrule
            B0 & Base no domain-id & 39.9 (30.1/1.2/8.6) & 30 \\
            B1-c & B1 caption-id & 30.6 (17.9/1.5/11.1) & 25 \\
            B1-d & B1 dictation-id & 68.7 (64.8/0.4/3.4) & 62 \\
            E1 & speaker-tags model & 37.5 (27.0/1.4/9.1) & 28 \\
        \bottomrule
    \end{tabular}
    \label{tab:del-res}
\end{table}

Domain-id input is critical to serve different applications with good WER quality (Table \ref{tab:dom-id}), however it makes the long-form deletion issue more pronounced (Table \ref{tab:del-res}). Ideally, we can train a model with no additional domain-id input for better cross-domain learning without hurting the WER performance on any domain. By unifying the output format for ASR with speaker-tags, this is now feasible. E1 in Table \ref{tab:dom-id} and Table \ref{tab:del-res} shows the performance of a model trained with speaker-tags and without domain-id as as input. Comparing E1 to the baseline B1, the WER on Short, Dictation and Caption is neutral or has very little regression. But E1 has substantially less long-form deletions than B1-d. The speaker-tags model alleviates the long-form deletion problem significantly.

These results indicate that the long-form deletion is not a problem when the ASR model is only trained with data from the Caption domain (similar to baseline B1-c in Table \ref{tab:del-res}). It becomes a problem when the ASR model is trained with data from other domains (similar to baseline B1-d in Table \ref{tab:del-res}). In domains like Short and Dictation, the ground truth is annotated in a way such that all non-primary speaker signals are ignored on purpose, and tends to end early; while in the Caption domain, the ground truth annotation attempts to cover all speech signals. The conflicting goals from different domains caused confusions during the model training, thus leading to long-form deletion errors when non-primary speakers are present. By explicitly modeling the primary and non-primary speakers in E1, we removed the confusion during the training, and fulfilled the domain-specific requirements as a post-processing step.

Next we examine the endpointer behavior (which is required only for Short domain). In Table \ref{tab:endpointer-res}, the EP50 and EP90 for E1 model with speaker-tags is about two times higher than the baseline. The model struggles to learn two conditions in the end - first emit a speaker-tag token then emit the \texttt{<end-of-speech>} token. To correct this, we assume that \texttt{<end-primary>} is also the endpointer signal. i.e. when the model emits the \texttt{<end-primary>} token, it's considered to be also emitting \texttt{<end-of-speech>} in parallel. E2 shows the results for such a model where the endpointer latency is significantly better and is on-par with the baseline. By merging the two tokens, the model is unable to handle cases where other speakers speak in the middle of the audio. During post processing, as soon as the first \texttt{<end-primary>} token is emitted, mic is closed. For example, if the ground truth is ``turn on the lights \texttt{<end-primary>} where is the book \texttt{<end-others>} in the bedroom \texttt{<end-primary>}", by the time the model outputs ``turn on the lights \texttt{<end-primary>}", the mic will be closed and the model will fail to transcribe ``in the bedroom". Although this is a limitation, such cases are very rare in Short and so, we see no WER degradation in practice. E2 has the same WER on Caption, Dictation (endpointer is needed only for Short) and the same long-form deletion performance as E1.

\begin{table}
    \centering
    \small
    \caption{Endpointer results. Note: unlike Table \ref{tab:dom-id}, here endpointer is enabled so WER is different for Short.}
    \begin{tabular}{c c c c c}
        \toprule
        \textbf{Exp} & \textbf{Setup} & \textbf{Short} & \textbf{EP50} & \textbf{EP90} \\
        \textbf{Id} & & \textbf{WER} & & \\
        \midrule
            B1 & Baseline & 6.2 & 260 & 830  \\
            E1 & speaker-tags model & 6.3 & 590 & 1625 \\
            E2 & E1 + merge tokens & 6.2 & 287 & 770 \\
        \bottomrule
    \end{tabular}
    \label{tab:endpointer-res}
\end{table}

Overall, E2 model with speaker-tags has substantially better long-form deletion performance compared to the baseline evaluated on dictation domain-id and performs almost as well as the baseline on all other testsets and metrics. Additionally, this model is capable of separating the output transcript by speaker groups and potentially unlocks new use cases (for instance, if a downstream application needs to switch between primary and non-primary speaker transcripts).

\section{Conclusions}

In this work, we proposed a technique to simultaneously model the primary and non-primary speakers along with the transcript tokens. This is achieved by introducing two new tokens into the wordpiece vocabulary, and relabeling the training data with two teacher models. The proposed approach enabled better cross-domain learning for large scale ASR training without additional training data. The improved cross domain learning reduces the training-test data mismatch, thereby significantly alleviating the long-form deletion problem. 

\section{Acknowledgements}
\ifblind
  Blinded for review
  
\else
 The authors would like to acknowledge the following people for many discussions that helped us shape this work: Weiran Wang, Arun Narayanan, Yanzhang He, Chao Zhang, Han Lu and Trevor Strohman, affiliated with Google LLC.
\fi

\bibliographystyle{IEEEbib}
\bibliography{refs}

\end{document}